\documentclass[
 prd,preprint,longbibliography,
 showpacs,showkeys,lengthcheck,
 nofootinbib,tightenlines,onecolumn,notitlepage,
 preprintnumbers,superscriptaddress
]{revtex4-1}

\usepackage[utf8]{inputenc}
\usepackage{newtxtext,newtxmath}
\usepackage[mathcal]{euscript}

\usepackage{graphicx}
\usepackage[dvipsnames]{xcolor}
\graphicspath{{figures/}}
\usepackage{tikz-feynman}
\usetikzlibrary{decorations.pathreplacing,calligraphy}
\usepackage{hyperref}
\hypersetup{colorlinks=true,citecolor=blue,linkcolor=blue,urlcolor=blue}
\usepackage{orcidlink}

\usepackage{diagbox}
\usepackage{booktabs}
\usepackage{siunitx}
\AtBeginDocument{%
  \heavyrulewidth=.08em
  \lightrulewidth=.05em
  \cmidrulewidth=.03em
  \belowrulesep=.65ex
  \belowbottomsep=0pt
  \aboverulesep=.4ex
  \abovetopsep=0pt
  \cmidrulesep=\doublerulesep
  \cmidrulekern=.5em
  \defaultaddspace=.5em
}

\usepackage{amsmath}
\usepackage{slashed}
\usepackage{bm}
\usepackage{bbm}
\usepackage{mathtools}
\usepackage{tensor}

\usepackage[math]{cellspace}
\usepackage[draft,commentmarkup=uwave]{changes} 
\definechangesauthor[color=green!50!black]{Adam}
\definechangesauthor[color=blue!70!black]{Jerry}

\renewcommand*{\d}{\mathop{}\!\mathrm{d}}
\newcommand*{\e}{\mathop{}\!\mathrm{e}}

\newcommand*{\nl}{n_{z\!}}
\newcommand*{\nt}{n_{\!\perp\!}}
\newcommand*{\ml}{m_l}

\usepackage{soul} 

\def\p{\perp}

\def\n{\nu}

\newcommand{\eq}[1]{Eq.~(\ref{#1})}

\def\n_p{n_\parallel}

\begin{document}

\title{
  Three-dimensional, boost-invariant formalism
  for systems of relativistically moving constituents
}

\author{Adam J. Freese \orcidlink{0000-0002-0688-4121}\,}
\email{afreese@jlab.org}
\affiliation{Center for Nuclear Femtography, Southeastern Universities Research Association, Newport News, Virginia 23606, USA}
\affiliation{Theory Center, Jefferson Lab, Newport News, Virginia 23606, USA}

\author{Anne L. Lashbrook \orcidlink{0009-0002-9120-7738}\,} 
\email{alash@uw.edu}
\affiliation{Department of Physics, University of Washington, Seattle, WA 98195-1560, USA}

\author{Gerald A. Miller \orcidlink{0000-0003-2443-3639\,}}
\email{miller@uw.edu}
\affiliation{Department of Physics, University of Washington, Seattle, WA 98195-1560, USA}

\begin{abstract}
  Light front quantum mechanics in three dimensions can be used to construct
  boost-invariant wave functions
  for the internal structure of relativistic systems.
  The Miller-Brodsky variable $\tilde{z}$---which is canonically conjugate
  to the momentum fraction $x$---allows a spatial description of
  the longitudinal degree of freedom.
  We show how $\tilde{z}$ can be constructed as an operator and prove
  its boost invariance.
  A relativistic harmonic oscillator potential
  from Li, Maris, Zhao and Vary~\cite{Li:2015zda}
  is used as an example of a two-body interaction
  that can be constructed using $\tilde{z}$
  and for which closed-form analytic solutions can be found.
  We systematically explore the conditions in which the non-relativistic
  harmonic oscillator solutions are reproduced
  and the conditions in which relativistic corrections are significant.
  Harmonic oscillator states are commonly used as a basis for nuclear many-body calculations.
  The present effort may provide a basis for providing light-front wave functions of nuclei.
\end{abstract}

\preprint{JLAB-THY-26-4829}
\preprint{NT@UW-26-17}

\maketitle


\section{Introduction}
\label{sec:intro}

Light front quantum mechanics and quantum field
theory~\cite{Dirac:1949cp,Kogut:1969xa,Soper:1971sr,Kogut:1972di,Brodsky:1997de,Miller:2000kv,Burkardt:2002hr,Vary:2009gt}
are the basis for most current studies
of hadronic structure when relativistic speeds are involved.
Other frameworks are in use, but have comparatively limited domains of applicability.
Non-relativistic quantum chromodynamics~\cite{Lucha:1991vn,Brambilla:1999xf,Brambilla:2004jw}
is successful for describing bound systems of heavy quarks,
but light front quantum mechanics is necessary to describe bound states of
light or massless quarks moving at nearly the speed of light.
Covariant bound state equations such as
the Bethe-Salpeter equation in principle allow manifestly covariant calculations
in momentum space without defining an equal-time surface,
but in practice the Bethe-Salpeter amplitude is usually projected
onto the light front to make solution
practical~\cite{Carbonell:1998rj,Sales:1999ec,Brodsky:2003pw,Cheng:2003sm,Karmanov:2005nv,Frederico:2011ws}\footnote{
  There are limited exceptions,
  which require restrictively simple interactions
  such as the contact interaction of the
  Nambu--Jona-Lasinio model~\cite{Nambu:1961tp,Nambu:1961fr};
  see Refs.~\cite{Cloet:2014rja,Freese:2019bhb}
  for examples.
}.

A framework as widely-used and successful as light front quantum mechanics
merits a pedagogical problem,
analogous to the hydrogen atom in non-relativistic quantum mechanics.
This problem would take the form of a wave equation for a composite system
of multiple particles bound by a specific realistic interaction
with an exact closed-form solution.
The solution would be boost-invariant by virtue of the light front's
Galilei subgroup~\cite{Dirac:1949cp,Susskind:1967rg,Kogut:1969xa,Soper:1971sr,Burkardt:2002hr,Lorce:2018zpf},
which uniquely allows separation of variables between barycentric
and internal degrees of freedom~\cite{Brodsky:1997de,Brodsky:2006uqa,Vary:2009gt}.
Such a problem would not only serve students learning
light front quantum mechanics,
but could also be used as a simple case in studies
of the various distribution functions
and form factors being used to describe hadron structure in the literature.

In this work, we consider the light front harmonic oscillator
as an outstanding candidate for such a problem.
The exact form of the harmonic oscillator potential we use
was previously postulated by
Li, Maris, Zhao and Vary (LMZV)~\cite{Li:2015zda}.
In this work, however, we go beyond simply postulating and solving a wave equation.
We show how the LMVZ potential is naturally expressed in terms of the
Miller-Brodsky variable~\cite{Miller:2019ysh},
which provides a boost-invariant description of longitudinal spatial distances.
The light front harmonic oscillator thus also provides an opportunity
for exploring the relationship between
spatial and momentum descriptions of bound state structure
in the light front formalism.

We stress that it is the separation of variables that makes light front
unique for calculating bound state structure.
While some prior works~\cite{Blunden:1999hy,Blunden:1999gq,Smith:2002ci}
found approximate solutions of the Dirac equation in light front coordinates
for a single particle under the influence of an external field,
this is also possible using instant form coordinates;
Darwin~\cite{darwin1928dirac}
solved the Dirac equation for an electron in a Coulomb field in 1928.
By contrast, attempts to formulate bound state equations
for two particles in instant form coordinates have been less successful.
For example, Feynman, Kislinger and Ravndal~\cite{Feynman:1971wr}
found the excitation of timelike modes for the harmonic oscillator to give troubles,
which they circumvented by arbitrarily removing the troublesome modes.
Leutwyler and Stern~\cite{Leutwyler:1977vy}
encountered the same problem,
which they avoided the same way.
The light front harmonic oscillator---using the LMZV potential---does
not encounter such troubles, and is exactly solvable.

This work is organized as follows.
Sec.~\ref{sec:lf} gives an overview of light front coordinates.
This includes the construction of light front position operators
in Sec.~\ref{sec:operators},
separation between barycentric and internal variables
in Sec.~\ref{sec:separation},
the definition and operator construction of
Miller-Brodsky variable in Sec.~\ref{sec:miller},
and a brief review of de Teramond's and Brodsky's
holographic variable~\cite{Brodsky:2006uqa,Brodsky:2014yha} in Sec.~\ref{sec:holo}.
After this, Sec.~\ref{sec:harmonic} uses the formalism of the previous section
to pose (in Sec.~\ref{sec:potential}) and solve (in Sec.~\ref{sec:solutions})
the light front harmonic oscillator problem.
We also explore the non-relativistic limit in Sec.~\ref{sec:nr},
showing that the correct solutions are reproduced
in the non-relativistic limit.
We lastly conclude and provide an outlook in Sec.~\ref{sec:end}.


\section{Light front coordinates}
\label{sec:lf}

Light front coordinates
reparametrize spacetime using the definitions:
\begin{align}
  \begin{split}
    x^\pm
    &=
    \frac{1}{\sqrt{2}}
    \big( t \pm z \big)
    \\
    \bm{x}_\perp
    &=
    (x,y)
    \,,
  \end{split}
\end{align}
with $x^+ = \frac{1}{\sqrt{2}}\big(t+z\big)$
chosen to be the time variable.
Operationally, using light front coordinates means taking what an observer
looking in the $+z$ direction sees at face value:
whatever they see at present is considered to be happening at present;
transverse distortions are formulated as Terrell rotations~\cite{Terrell:1959zz}
instead of Fitzgerald length contraction;
and moving clocks are described in terms of redshift and blueshift
formulas rather than a symmetric time dilation.
Formally, light front coordinates benefit from a large subgroup of the Poincar\'e group
that leaves the foliation of spacetime into fixed-$x^+$ surfaces intact,
and which therefore maintain the absoluteness of light front simultaneity.
Crucially, this subgroup includes boosts,
which allows internal structure of relativistic bound states to be
disentangled from the composite system's barycentric motion.
See Refs.~\cite{Dirac:1949cp,Soper:1971sr,Kogut:1972di,Brodsky:1997de,Miller:2000kv,Burkardt:2002hr} for reviews,
Ref.~\cite{Freese:2023jcp} for a more conceptual overview,
and Appendix~\ref{sec:poincare} for a brief technical overview.

Four-momenta are, as usual,
defined to be the generators of spacetime translations:
\begin{align}
  p_\mu
  =
  i \partial_\mu
  \,,
\end{align}
and $p^- = p_+ = i \partial_+$
takes the role of a Hamiltonian operator that generates time translations
(i.e., translations in $x^+$).
The operator $p^+$ generates longitudinal translations in $x^-$,
and accordingly is the longitudinal momentum operator.
The light front kinetic energy
is taken from the
relativistic mass-shell condition:
\begin{align}
  m^2
  =
  2 p^+ p^-
  -
  \bm{p}_\perp^2
\end{align}
so that the
formula for the kinetic energy is given by
\begin{align}
  \label{eqn:H:free}
  H  =  p^-  = \frac{m^2 + \bm{p}_\perp^2}{2p^+}
\end{align}
This is
reminiscent of non-relativistic formula for kinetic energy~\cite{Kogut:1972di} in two spatial dimensions.


\subsection{Light front position operators}
\label{sec:operators}

The coordinate operator
$\bm{X}_\perp$ and the momentum operator $\bm{P}_\perp$%
---which we signify with capital letters in this section
to distinguish from c-numbers---%
are canonically conjugate,
and obey the standard commutation relation
\begin{align}
  [X_\perp^i, P_\perp^j]
  =
  i
  \delta^{ij}
  \,.
\end{align}
Just as in non-relativistic quantum mechanics,
one can use $\bm{P}_\perp = -i \bm{\nabla}^{(x)}_\perp$ in the coordinate representation
and $\bm{X}_\perp = +i \bm{\nabla}^{(p)}_\perp$ in the momentum representation.
The longitudinal position and momentum operators should obey similar commutation relations, namely:
\begin{align}
  \begin{split}
    [X^-, P^+]
    &=
    -i
    \\
    [X^-, P_\perp^i]
    &=
    0
    \\
    [X_\perp^i, P^+]
    &=
    0
    \,,
  \end{split}
\end{align}
but require a different treatment than $\bm{X}_\perp$ and $\bm{P}_\perp$
because of the relativistic normalization convention.
In the coordinate space, one has
$P^+ = i \partial^+ = i\partial_-$,
but in the momentum representation we instead have:
\begin{align}
  \label{eqn:xminus}
  \langle p^+, \bm{p}_\perp |
  X^-
  | \psi \rangle
  =
  -i
  \left(
  \frac{\partial}{\partial p^+}
  -
  \frac{1}{2p^+}
  \right)
  \Big[
    \langle p^+, \bm{p}_\perp | \psi \rangle
    \Big]
  \,,
\end{align}
somewhat akin to the Newton-Wigner operator~\cite{Newton:1949cq} in instant form.
This form is imposed by the requirement that $X^-$ be Hermitian:
\begin{align}
  \label{eqn:hermiticity}
  \langle \phi | X^- | \psi \rangle
  =
  \langle \phi | (X^-)^\dagger | \psi \rangle
  =
  \big(\langle \psi | X^- | \phi \rangle\big)^*
  \,,
\end{align}
together with the Lorentz-invariant normalization rule for one-particle states:
\begin{align}
  \label{eqn:completeness}
  \int \frac{\d p^+ \d^2 p_\perp}{2p^+(2\pi)^3}
  | p^+, \bm{p}_\perp \rangle \langle p^+, \bm{p}_\perp |
  =
  1
  \,.
\end{align}
Inserting Eq.~(\ref{eqn:completeness})
into the far-left and far-right sides of Eq.~(\ref{eqn:hermiticity}) gives:
\begin{align}
  \int \frac{\d p^+ \d^2 p_\perp}{2p^+(2\pi)^3}
  \langle \phi | p^+, \bm{p}_\perp \rangle
  \langle p^+, \bm{p}_\perp |X^- | \psi \rangle
  =
  \int \frac{\d p^+ \d^2 p_\perp}{2p^+(2\pi)^3}
  \big( \langle \psi | p^+, \bm{p}_\perp \rangle \big)^*
  \big( \langle p^+, \bm{p}_\perp |X^- | \phi \rangle \big)^*
  \,.
\end{align}
Using Eq.~(\ref{eqn:xminus}) on both sides of this equation,
as well as the identity $(\langle A | B \rangle)^* = \langle B | A \rangle$,
in turn gives:
\begin{align}
  -i
  \int \frac{\d p^+ \d^2 p_\perp}{2p^+(2\pi)^3}
  \langle \phi | p^+, \bm{p}_\perp \rangle
  \left(
  \frac{\partial}{\partial p^+}
  -
  \frac{1}{2p^+}
  \right)
  \Big[
    \langle p^+, \bm{p}_\perp | \psi \rangle
    \Big]
  =
  i
  \int \frac{\d p^+ \d^2 p_\perp}{2p^+(2\pi)^3}
  \langle p^+, \bm{p}_\perp | \psi \rangle
  \left(
  \frac{\partial}{\partial p^+}
  -
  \frac{1}{2p^+}
  \right)
  \Big[
    \langle \phi | p^+, \bm{p}_\perp \rangle
    \Big]
  \,.
\end{align}
This result can be confirmed using integration by parts.
Notably, when using integration by parts,
the $p^+$ derivative ends up acting on the factor
$\frac{1}{p^+}$ in the integration measure,
and the presence of $-\frac{1}{2p^+}$ in $X^-$
thus becomes necessary for Hermiticity to work out.

\begin{table}
  \renewcommand{\arraystretch}{2.0}
  \caption{
    Table of light front position and momentum operators
    in the coordinate, momentum and mixed representations.
  }
  \begin{tabular}{ccccc}
    \toprule
    ~ &
    ~~~~$\bm{X}_\perp$~~~~ &
    ~~~~$\bm{P}_\perp$~~~~ &
    ~~~~$X^-$~~~~ &
    ~~~~$P^+$~~~~ \\
    \hline
    Position representation &
    $\bm{x}_\perp$ &
    $-i\bm{\nabla}^{(x)}_\perp$ &
    $x^-$ &
    $i\partial_-$ \\
    \hline
    Momentum representation &
    $i\bm{\nabla}^{(p)}_\perp$ &
    $\bm{p}_\perp$ &
    $-i\left(\frac{\partial}{\partial p^+} - \frac{1}{2p^+}\right)$ &
    $p^+$ \\
    \hline
    Mixed representation &
    $\bm{x}_\perp$ &
    $-i\bm{\nabla}^{(x)}_\perp$ &
    $-i\left(\frac{\partial}{\partial p^+} - \frac{1}{2p^+}\right)$ &
    $p^+$ \\
    \bottomrule
  \end{tabular}
  \label{tab:operators}
\end{table}

Besides the coordinate and momentum representations,
it is also helpful in light front coordinates to consider a
mixed representation of transverse position and longitudinal momentum,
for which the wave equation takes the form:
\begin{align}
  i \partial_+
  \psi(p^+,\bm{x}_\perp)
  =
  \frac{m^2 - \bm{\nabla}_\perp^2}{2p^+}
  \psi(p^+,\bm{x}_\perp)
  \,.
\end{align}
The mixed representation wave function is related to the momentum-space
wave function through a two-dimensional Fourier transform:
\begin{align}
  \psi(p^+,\bm{x}_\perp)
  &=
  \int \frac{\d^2 p_\perp}{(2\pi)^2}
  \psi(p^+, \bm{p}_\perp)
  \e^{i\bm{p}_\perp\cdot\bm{x}_\perp}
  \,,
\end{align}
and accordingly obeys the normalization rule:
\begin{align}
  \int \frac{\d p^+ \d^2x_\perp}{4\pi p^+}
  \big|
  \psi(p^+,\bm{x}_\perp)
  \big|^2
  =
  1
  \,.
\end{align}

The forms of the light front position and momentum operators in
the position, momentum and mixed representations are all
given in Table~\ref{tab:operators}.


\subsection{Light front separation of variables}
\label{sec:separation}

In dealing with systems of composite particles,
it is necessary to separate barycentric and internal variables.
In the two-body problem,
the separation of variables for transverse components of position and momentum
proceeds similarly to the non-relativistic case:
\begin{align}
  \begin{split}
    \bm{R}_\perp
    =
    \frac{p^+_1 \bm{x}_{1\perp} + p^+_2 \bm{x}_{2\perp}}{p^+_1 + p^+_2}
    \,,
    & \qquad
    \bm{r}_\perp
    =
    \bm{x}_{1\perp} - \bm{x}_{2\perp}
    \,,
    \\
    \bm{P}_\perp
    =
    \bm{p}_{1\perp} + \bm{p}_{2\perp}
    \,,
    & \qquad
    \bm{k}_\perp
    =
    \frac{p^+_2 \bm{p}_{1\perp} - p^+_1 \bm{p}_{2\perp}}{p^+_1 + p^+_2}
    \,,
  \end{split}
\end{align}
but with $p^+$ appearing in place of the mass.
Note that we now use capital letters for barycentric degrees of freedom
and lowercase for constituent particles.
To perform separation of variables for the longitudinal components,
we must carefully consider the transformation properties
of the light front interaction Hamiltonian.
The full Hamiltonian is given by a sum of the free Hamiltonians---which
take the form (\ref{eqn:H:free})---and an interaction term:
\begin{align*}
  P^-
  =
  \frac{m_1^2 + \bm{p}_{1\perp}^2}{2p_1^+}
  +
  { \frac{m_2^2 + \bm{p}_{2\perp}^2}{2p_2^+}}
  +
  P^-_{\mathrm{int}}
  \,.
\end{align*}
The longitudinal variable dependence can be parametrized in terms of
a total longitudinal momentum $P^+$
and momentum fractions $x_1\equiv x$ and $x_2=1-x$:
\begin{align}
  P^+ = p_1^+ + p_2^+
  \,, \qquad
  x_1 = \frac{p_1^+}{P^+} \equiv x
  \,, \qquad
  x_2 = \frac{p_2^+}{P^+} = 1-x
  \,.
  \label{xidef}
\end{align}
The longitudinal momenta $p_1^+$ and $p_2^+$ are invariant under
the light front Galilei subgroup of the Poincar\'e group,
which includes light front transverse boosts~\cite{Susskind:1967rg,Kogut:1969xa,Soper:1971sr,Burkardt:2002hr,Lorce:2018zpf}.
Additionally, they scale in the same way under longitudinal boosts,
so the momentum fractions $x_1$ and $x_2$ are invariant under
all light front boosts.
This makes them especially useful to describing internal structure
in a manner that is independent of the barycentric state of motion,
and thus for achieving separation of variables.

A mass operator $M^2$ may be defined as
$M^2 \equiv 2 P^+ P^- - \bm{P}_\perp^2$.
This can be decomposed, like $P^-$,
into a free piece and an interaction piece:
\begin{align}
  \label{msq}
  \begin{split}
    M^2
    &=
    M^2_{\mathrm{free}}
    +
    M^2_{\mathrm{int}}
    \\
    M^2_{\mathrm{free}}
    &=
    \frac{m_1^2}{x}
    +
    \frac{m_2^2}{1-x}
    +
    \frac{\bm{k}_\perp^2}{x (1-x)}
    \\
    M^2_{\mathrm{int}}
    &=
    2P^+
    P^-_{\mathrm{int}}
    \,.
  \end{split}
\end{align}
By noting that $\bm{k}_\perp = -i \bm{\nabla}_\perp$
in the transverse coordinate representation,
the light front Sch\"odinger equation for
the internal structure of a two-body system can be written:
\begin{eqnarray}
  \label{eqn:schrodinger}
  -{\nabla_\perp^2\psi(x,\bm{r}_\p) \over x(1-x)}
  +
  \left({m_1^2\over x}+{m_2^2\over 1-x}\right)\psi(x,\bm{r}_\p)
  +
  M^2_{\mathrm{int}}
  \psi(x,\bm{r}_\p)
  =
  M^2
  \psi(x,\bm{r}_\p)
  \,.
\end{eqnarray}
The total $M^2$ is a Poincar\'e-invariant quantity.
Additionally, the momentum fractions $x_1$ and $x_2$,
as well as the relative transverse momentum $\bm{k}_\perp$,
are boost-invariant,
and thus independent of
the barycentric momentum $(P^+;\bm{P}_\perp)$.
This means the interaction term $M^2_{\mathrm{int}}$
is also independent of the barycentric momentum,
and accordingly represents the internal structure of the bound system.

Since $M^2_{\mathrm{int}}$ is boost-invariant,
it should be constructed from boost-invariant variables.
So far, we have
$x$, $\bm{k}_\perp$ and $\bm{r}_\perp$
as boost-invariant internal variables.
Conspicuously absent from this list is the longitudinal separation:
\begin{align*}
  z_1 - z_2
  =
  \left(\frac{x^+_1 - x^-_1}{\sqrt{2}}\right)
  -
  \left(\frac{x^+_2 - x^-_2}{\sqrt{2}}\right)
  \xrightarrow[\text{fixed $x^+$}]{}
  \left(\frac{x^-_2 - x^-_1}{\sqrt{2}}\right)
  \,.
\end{align*}
This operator does not commute with $\bm{R}_\perp$
(and is accordingly not independent of the barycentric state),
and is not invariant under longitudinal or transverse boosts.
However, the Miller-Brodsky variable~\cite{Miller:2019ysh}
\begin{align}
  \label{eqn:MB}
  \tilde{z}
  \equiv
  P^+
  \big( x_2^- - x_1^- \big)
  +
  \bm{P}_\perp \cdot \bm{r}_\perp
\end{align}
is both boost-invariant and independent
of the barycentric state\footnote{
  Strictly speaking, Ref.~\cite{Miller:2019ysh}
  assumed $\bm{P}_\perp=0$,
  so Eq.~(\ref{eqn:MB}) here is a boost-invariant generalization.
}.
It is additionally the canonically conjugate variable to
the momentum fraction $x$.
(A proof of these statements is contained in Appendix~\ref{sec:poincare}.)
The Miller-Brodsky variable can accordingly be added
to the list of good variables for constructing $M_{\mathrm{int}}^2$.

The quantities $\bm{k}_\perp$ and $\bm{r}_\perp$ are canonically conjugate, and we
can express $\bm{k}_\perp $ as $ -i\bm{\nabla}_\perp$ in coordinate representation.
We need a similar relationship between
$x$ and $\tilde{z}$ to obtain a solvable wave equation.
From Eq.~(\ref{xidef}), in terms of the particle variables:
\begin{align}
  x_1^-
  =
  -i\left(\frac{\partial}{\partial p_1^+} - \frac{1}{2p_1^+}\right)
  \,,
  \qquad \qquad
  x_2^-
  =
  -i\left(\frac{\partial}{\partial p_2^+} - \frac{1}{2p_2^+}\right)
  \,.
\end{align}
When transforming to the barycentric and relative variables,
we must exploit the standard rule for transforming derivatives.
In this context, we have:
\begin{align}
  \begin{split}
    \frac{\partial}{\partial p^+_1}
    &=
    \frac{\partial x}{\partial p^+_1}
    \frac{\partial}{\partial x}
    +
    \frac{\partial P^+}{\partial p^+_1}
    \frac{\partial}{\partial P^+}
    +
    \frac{\partial R_\perp^a}{\partial p^+_1}
    \frac{\partial}{\partial R_\perp^a}
    =
    \frac{1-x}{P^+}
    \frac{\partial}{\partial x}
    +
    \frac{\partial}{\partial P^+}
    +
    \frac{\bm{x}_{1\perp} - \bm{R}_\perp}{P^+}
    \cdot
    \bm{\nabla}_\perp^{(R)}
    \\
    \frac{\partial}{\partial p^+_2}
    &=
    \frac{\partial x}{\partial p^+_2}
    \frac{\partial}{\partial x}
    +
    \frac{\partial P^+}{\partial p^+_2}
    \frac{\partial}{\partial P^+}
    +
    \frac{\partial R_\perp^a}{\partial p^+_2}
    \frac{\partial}{\partial R_\perp^a}
    =
    -\frac{x}{P^+}
    \frac{\partial}{\partial x}
    +
    \frac{\partial}{\partial P^+}
    +
    \frac{\bm{x}_{2\perp} - \bm{R}_\perp}{P^+}
    \cdot
    \bm{\nabla}_\perp^{(R)}
    \,,
  \end{split}
\end{align}
and thus:
\begin{align}
  x_2^- - x_1^-
  =
  i
  \left(
  \frac{\partial}{\partial x}
  -
  \frac{1-2x}{2x(1-x)}
  +
  \bm{r}_\perp \cdot \bm{\nabla}_\perp^{(R)}
  \right)
  \,.
\end{align}
This allows the action of the Miller-Brodsky coordinate
in momentum fraction space to be written entirely
in terms of $x$ and its derivatives:
\begin{align}
  \label{eqn:ztilde}
  \tilde{z} \psi(x,\bm{r}_\perp)
  =
  \Big\{
    P^+ \big( x_2^- - x_1^- \big)
    +
    \bm{P}_\perp\cdot\bm{r}_\perp
    \Big\}
  \psi(x,\bm{r}_\perp)
  =
  i
  \sqrt{x(1-x)}
  \frac{\partial}{\partial x}
  \left[
    \frac{1}{\sqrt{x(1-x)}}
    \psi(x,\bm{r}_\perp)
    \right]
  \,.
\end{align}
A summary of the good, boost-invariant quantities for building $M^2_{\mathrm{int}}$
and their actions on the mixed-representation wave function $\psi(x,\bm{r}_\perp)$
are given in Table~\ref{tab:good}.

\begin{table}
  \renewcommand{\arraystretch}{2.0}
  \caption{
    Table of good, boost-invariant variables for describing internal
    structure, and their action on the internal wave function
    $\psi(x,\bm{r}_\perp)$.
  }
  \begin{tabular}{ccccc}
    \toprule
    ~ &
    ~~~~~~$\bm{r}_\perp$~~~~~~ &
    ~~~~~~$\bm{k}_\perp$~~~~~~ &
    ~~~~~~$x$~~~~~~ &
    ~~~~~~$\tilde{z}$~~~~~~ \\
    \hline
    Action on $\psi(x,\bm{r}_\perp)$~~ &
    ~~$\bm{r}_\perp\psi(x,\bm{r}_\perp)$~~ &
    ~~$-i\bm{\nabla}^{(r)}_\perp\psi(x,\bm{r}_\perp)$~~ &
    ~~$x\psi(x,\bm{r}_\perp)$~~ &
    ~~$
    i
    \sqrt{x(1-x)}
    \frac{\partial}{\partial x}
    \left[
      \frac{1}{\sqrt{x(1-x)}}
      \psi(x,\bm{r}_\perp)
      \right]
    $~~
    \\
    \bottomrule
  \end{tabular}
  \label{tab:good}
\end{table}

We lastly turn to the normalization of the wave function.
Since $M^2$ depends only on internal variables,
the relative wave function can be normalized in a manner dependent on $x$:
\begin{align}
  \int \frac{\d x}{4\pi x(1-x)}
  \int \frac{\d^2 k_\perp}{(2\pi)^2} \,
  \big| \psi(x,\bm{k}_\perp) \big|^2
  =
  \int \frac{\d x}{4\pi x(1-x)}
  \int \d^2r_\perp \,
  \big| \psi(x,\bm{r}_\perp) \big|^2
  =
  1
  \,.
  \label{norm}
\end{align}
Most significantly,
the presence of $\frac{1}{x(1-x)}$ in the integration element ensures that
$\tilde{z}$---as defined in Eq.~(\ref{eqn:ztilde})---is Hermitian.
This can be proved in a similar manner to the derivation of Eq.~(\ref{eqn:xminus}).


\subsection{Wave function in Miller-Brodsky coordinate representation}
\label{sec:miller}

While it is easiest to work in the mixed $(x,\bm{r}_\perp)$ representation
for the internal structure of two-body systems,
it can be enlightening---once a solution is obtained---to transform to the
$(\tilde{z}, \bm{r}_\perp)$ representation to examine the longitudinal
spatial structure of a system.

We choose the $(\tilde{z}, \bm{r}_\perp)$ representation wave function
to be normalized as:
\begin{align}
  \label{eqn:norm:mb}
  \int_{-\infty}^{\infty} \d \tilde{z}
  \int \d^2 r_\perp \,
  \big|
  \psi(\tilde{z}, \bm{r}_\perp)
  \big|^2
  =
  1
  \,.
\end{align}
The wave function in this representation is related to the
$(x,\bm{r}_\perp)$ representation wave function through a Fourier transform:
\begin{align}
  \psi(\tilde{z}, \bm{r}_\perp)
  =
  \int_0^1 \d x \mu(x) \,
  \psi(x, \bm{r}_\perp)
  \e^{ix\tilde{z}}
  \,,
\end{align}
where the real-valued measure function $\mu(x)$
is determined by the joint requirements of
Eq.~(\ref{norm}) and Eq.~(\ref{eqn:norm:mb}).
One can thus show that:
\begin{align}
  \psi(\tilde{z}, \bm{r}_\perp)
  =
  \int_0^1 \d x \,
  \frac{
    \psi(x, \bm{r}_\perp)
  }{
    2\pi \sqrt{2x(1-x)}
  }
  \e^{ix\tilde{z}}
  \,.
\end{align}
The presence of $\frac{1}{\sqrt{x(1-x)}}$ may look peculiar,
but is correct.
Since $x$ is boost-invariant,
the integration measure is also boost-invariant.
Additionally,
the presence of this factor ensures that $\tilde{z}$ as an operator
behaves as expected in both the $(\tilde{z},\bm{r}_\perp)$
and the $(x,\bm{r}_\perp)$ representations.
In fact, through integration by parts, one can show:
\begin{align}
  \tilde{z}
  \psi(\tilde{z}, \bm{r}_\perp)
  =
  \int_0^1 \d x \,
  \frac{
    1
  }{
    2\pi \sqrt{2x(1-x)}
  }
  \left\{
    i
    \sqrt{x(1-x)}
    \frac{\partial}{\partial x}
    \left[
      \frac{1}{\sqrt{x(1-x)}}
      \psi(x, \bm{r}_\perp)
      \right]
    \right\}
  \e^{ix\tilde{z}}
  \,.
\end{align}
Here, the quantity in the curly brackets $\{\,\}$
is consistent with the rule for the action of the Miller-Brodsky variable
in the $(x,\bm{r}_\perp)$ representation;
see Eq.~(\ref{eqn:ztilde}).


\subsection{Transformation to holographic variables}
\label{sec:holo}

Another helpful---and very commonly-used---variable transformation
to consider is the Brodsky-de Teramond holographic
variable~\cite{Brodsky:2006uqa,Brodsky:2014yha}:
\begin{eqnarray}
  \label{eqn:zeta}
  \zeta
  \equiv
  \sqrt{x(1-x)}
  \,
  r_\perp
  \,.
\end{eqnarray}
When performing a coordinate transformation from
the primitive variables $(x, r_\perp, \phi)$
to the holographic variables $(x', \zeta', \phi')$:
\begin{align}
  \begin{split}
    x' &= x
    \\
    \zeta' &= \sqrt{x(1-x)} r_\perp
    \\
    \phi' &= \phi
  \end{split}
\end{align}
we must use the standard rule for transformations of derivatives:
\begin{align}
  \frac{\partial}{\partial x^a}
  =
  \frac{\partial x'^b}{\partial x^a}
  \frac{\partial}{\partial x'^b}
  \,.
\end{align}
Although $(x',\phi') = (x,\phi)$, the temporary use of primes will help us
avoid mistakes, by reminding us that we are transforming to a new coordinate system.
Most crucially,
\begin{align}
  \frac{\partial \zeta'}{\partial x}
  =
  \frac{1-2x'}{2x'(1-x')} \zeta'
  \neq
  0
  \,,
\end{align}
and thus partial $x$ derivatives at fixed $r_\perp$ and fixed $\zeta$ are not the same.
This happens because $r_\perp$ and $\zeta$ cannot be simultaneously held fixed
while $x$ is varied.
In particular:
\begin{align}
  \label{eqn:tricky}
  \begin{split}
    \frac{\partial}{\partial x}\bigg|_{r}
    &=
    \frac{\partial}{\partial x}\bigg|_{\zeta}
    +
    \frac{1-2x}{2x(1-x)}
    \zeta
    \frac{\partial}{\partial \zeta}
    \\
    \frac{\partial}{\partial r_\perp}
    &=
    \sqrt{x(1-x)}
    \frac{\partial}{\partial \zeta}
    \\
    \frac{\partial}{\partial \phi}
    &=
    \frac{\partial}{\partial \phi}
    \,.
  \end{split}
\end{align}
Here, instead of using primes,
we follow the common notation (e.g., from thermodynamics texts) that
$\frac{\partial}{\partial x}\big|_{r}$ means a partial derivative
with respect to $x$ while $r$ is held fixed.

The original motivation---and the namesake---for the holographic variable
comes from a formal correspondence between light front descriptions of hadron
structure and field theories in anti-deSitter spacetime,
as explored by Brodsky, de Teramond and
others~\cite{Brodsky:2006uqa,Brodsky:2014yha,Li:2015zda,Miller:2019ysh,Li:2021jqb}.
Our primary motivation to adopt it is instead grounded in its utility
for separation of variables,
as will be explored in Sec.~\ref{sec:harmonic}
with the light front harmonic oscillator as a concrete example.
Hints of its utility can be seen already by rewriting
the light front Schr\"odinger equation (\ref{eqn:schrodinger})
in terms of holographic variables:
\begin{align}
  \label{eqn:schrodinger:holo}
  \left(
  -
  {\partial^2 \over \partial \zeta^2}
  -
  {1\over \zeta}{\partial \over \partial \zeta}
  -
  {1\over \zeta^2}{\partial^2\over \partial \phi^2}
  +
  {m_1^2\over x}
  +
  {m_2^2\over 1-x}
  +
  M_{\mathrm{int}}^2
  \right)
  \psi(x,\zeta,\phi)
  =
  M^2
  \psi(x,\zeta,\phi)
  \,.
\end{align}
The $(\zeta,\phi)$ derivatives now have no factors with $x$ dependence.
Whether separation of variables can be achieved now depends only on
the form of the interaction Hamiltonian $M_{\mathrm{int}}^2$.

A transformation to holographic variables will transform the integration
measure and normalization rule for the light front wave function.
The Jacobian for the transformation gives:
\begin{align}
  \label{eqn:norm:holo}
  \int_0^1 \frac{\d x}{4\pi x^2(1-x)^2}
  \int_0^\infty \d \zeta \, \zeta
  \int_0^{2\pi} \d \phi \,
  \big| \psi(x,\zeta,\phi) \big|^2
  =
  1
  \,.
\end{align}

Transforming to holographic variables also requires modifying
the Miller-Brodsky variable.
$\tilde{z}$ commutes with $r_\perp$
and has the correct canonical commutation rule with $x$,
but it does not commute with $\zeta$:
\begin{align}
  [ \tilde{z}, \zeta ]
  =
  -
  \frac{i(1-2x)}{2x(1-x)}
  \zeta
  \neq
  0
  \,.
\end{align}
The essence of the issue is precisely that $\tilde{z}$ involves
a partial $x$ derivative with $r_\perp$ rather than $\zeta$ held fixed.
However, simply replacing
$\frac{\partial}{\partial x}\big|_{r_\perp}$
with
$\frac{\partial}{\partial x}\big|_{\zeta}$
in Eq.~(\ref{eqn:ztilde})
will produce an operator that is not Hermitian;
the change to holographic variables has changed the integration measure,
as in Eq.~(\ref{eqn:norm:holo}),
and the new Miller-Brodsky variable must be Hermitian with respect
to the new measure.
Besides replacing
$\frac{\partial}{\partial x}\big|_{r_\perp}
\rightarrow
\frac{\partial}{\partial x}\big|_{\zeta}$,
one must also replace $\sqrt{x(1-x)} \rightarrow x(1-x)$
in Eq.~(\ref{eqn:ztilde})---that is, one must use the square root
of the new integration measure to ensure Hermiticity---giving:
\begin{align}
  \label{eqn:wtilde}
  \tilde{z}_{\zeta}
  \psi(x, \zeta, \phi)
  =
  i x(1-x)
  \frac{\partial}{\partial x}
  \left[
    \frac{ \psi(x, \zeta, \phi) }{x(1-x)}
    \right]
  \bigg|_{\zeta}
  \,.
\end{align}
We will call this the \emph{holographic} Miller-Brodsky variable,
and use a subscript $\zeta$
to distinguish it from the standard Miller-Brodsky variable.

With a bit of effort,
one can rewrite the holographic Miller-Brodsky variable
in terms of the $\bm{r}_\perp$ and $\bm{k}_\perp$ operators as:
\begin{align}
  \tilde{z}_\zeta
  =
  \tilde{z}
  +
  \frac{1-2x}{2x(1-x)}
  \frac{\bm{k}_\perp\cdot\bm{r}_\perp + \bm{r}_\perp\cdot\bm{k}_\perp}{2}
  \,,
\end{align}
and show that it acts on a wave function in the $(x,\bm{r}_\perp)$
representation in the following way:
\begin{align}
  \label{eqn:complicated}
  \tilde{z}_{\zeta}
  \psi(x, \bm{r}_\perp)
  =
  i
  \left(
  \frac{\partial}{\partial x}\bigg|_{r_\perp}
  -
  \frac{1-2x}{x(1-x)}
  -
  \frac{1-2x}{2x(1-x)}
  r_\perp \frac{\partial}{\partial r_\perp}
  \right)
  \psi(x, \bm{r}_\perp)
  \,.
\end{align}
Since $\tilde{z}_\zeta$ depends only on the good internal variables
laid out in Table~\ref{tab:good}, it is also a good internal variable.


\section{Light front harmonic oscillator}
\label{sec:harmonic}

In the previous section,
we showed that the wave equation (\ref{msq})
could be written as a differential equation in terms of
the internal spatial variable $\bm{r}_\perp$ and momentum fraction $x$.
Here we provide the harmonic oscillator potential as a specific example
of a $M^2_{\mathrm{int}}$ for which closed-form analytic solutions can be obtained.
The harmonic oscillator is interesting not only as an especially well-behaved toy problem,
but is also practical as an approximation of various potentials at positions near equilibrium.
It not only has seen success in molecular and nuclear physics,
but two-dimensional harmonic oscillator potentials have even been successfully
used as a confining potential in effective models of quantum
chromodynamics~\cite{Brodsky:2007hb,Brodsky:2014yha,Li:2015zda,Li:2021jqb}.
The three-dimensional light front harmonic oscillator therefore has promise
as a three-dimensional extension of these successful models.


\subsection{Light front harmonic oscillator potential}
\label{sec:potential}

We choose a potential for the light front harmonic oscillator
that:
(a) is Hermitian;
(b) depends only on the good internal variables
laid out in Table~\ref{tab:good};
(c) reduces to the standard harmonic oscillator potential
in the non-relativistic limit; and
(d) has closed-form analytic solutions.
A potential that satisfies these requirements has been
proposed by Li, Maris, Zhao and Vary (LMZV)~\cite{Li:2015zda},
which we shall call the LMZV potential:
\begin{eqnarray}
  \label{eqn:ho}
  M^2_{\mathrm{int}}
  =
  \kappa^4
  \left(
  x(1-x) \bm{r}_\perp^2
  +
  \frac{
    \tilde{z}_\zeta x(1-x) \tilde{z}_\zeta
  }{
    (m_1+m_2)^2
  }
  \right)
  \,.
\end{eqnarray}
The potential is defined in terms of the holographic Miller-Brodsky variable,
because holographic variables are needed to achieve separation of variables.
It should be noted that by Eq.~(\ref{eqn:complicated}),
the potential will have a significantly more complicated form in the
physical $(x,\bm{r}_\perp)$ variables.

Now, from Eq.~(\ref{eqn:wtilde}),
the action of the potential in Eq.~(\ref{eqn:ho}) on the wave function is:
\begin{align}
  \label{eqn:Mint:holo}
  M^2_{\mathrm{int}}
  \psi(x,\zeta,\phi)
  =
  \kappa^4
  \left(
  \zeta^2
  \psi(x,\zeta,\phi)
  -
  \frac{x(1-x)}{(m_1+m_2)^2}
  \frac{\partial}{\partial x}
  \left[
    x(1-x)
    \frac{\partial}{\partial x}
    \left[
      \frac{\psi(x,\zeta,\phi)}{x(1-x)}
      \right]
    \right]
  \right)
  \,.
\end{align}
Thus, the holographic Schr\"odinger equation (\ref{eqn:schrodinger:holo})
for the harmonic oscillator potential (\ref{eqn:ho})
can be written in holographic variables as:
\begin{eqnarray}
  \label{eqn:holo}
  x(1-x)
  \left(
  -
  {\partial^2 \over \partial \zeta^2}
  -
  {1\over \zeta}{\partial \over \partial \zeta}
  -
  {1\over \zeta^2}{\partial^2\over \partial \phi^2}
  +
  {m_1^2\over x}
  +
  {m_2^2\over 1-x}
  -
  M^2
  +
  \kappa^4 \zeta^2
  -
  \frac{\kappa^4}{(m_1+m_2)^2}
  \frac{\partial}{\partial x}
  x(1-x)
  \frac{\partial}{\partial x}
  \right)
  \frac{
    \psi(x,\zeta,\phi)
  }{
    x(1-x)
  }
  =
  0
  \,.
\end{eqnarray}
This is amenable to solution using separation of variables.
The form of the potential in Eq.~(\ref{eqn:ho}) was crucial in ensuring
that this would be possible.


\subsection{Solutions to the light front harmonic oscillator}
\label{sec:solutions}

We will solve the Schr\"odinger equation (\ref{eqn:holo})
for the light front harmonic oscillator by separation of variables.
The result will be a collection of closed-form analytic wave functions
and eigenvalues for $M^2$ that are indexed by three quantum numbers
$(\nl,\nt,\ml)$.

To solve by separation of variables,
we factorize the wave function as:
\begin{eqnarray}
  \psi(x,\zeta,\phi)
  =
  x(1-x)
  X(x)
  Z(\zeta)
  \frac{\e^{i \ml \phi}}{\sqrt{2\pi}}
  \label{psi}
  \,.
\end{eqnarray}
Here, $\ml$ is the quantum number for the $z$ component of
the orbital angular momentum, and as such is an integer.
It coincides exactly with the standard $\ml$ quantum number
in non-relativistic quantum mechanics.
Note that the $\phi$ dependence obeys the orthonormality relation:
\begin{align}
  \label{eqn:ortho:phi}
  \frac{1}{2\pi}
  \int_0^{2\pi} \d\phi \,
  \e^{i \ml \phi}
  \e^{-i \ml' \phi}
  =
  \delta_{\ml \ml'}
  \,,
\end{align}
while the single-variable functions $X(x)$ and $Z(\zeta)$ satisfy
the normalization rules:
\begin{align}
  \label{eqn:norm:Z}
  \int_0^\infty \d \zeta \,
  \zeta
  |Z(\zeta)|^2
  &=
  1
  \\
  \label{eqn:norm:X}
  \int_0^1 \frac{\d x}{4\pi}
  |X(x)|^2
  &=
  1
  \,,
\end{align}
as follows from Eq.~(\ref{eqn:norm:holo}).
The Schr\"odinger equation (\ref{eqn:holo})
can now be decomposed into two ordinary differential equations:
\begin{align}
  -
  {\d^2Z(\zeta)\over \d\zeta^2}
  -
  {1\over\zeta}
  {\d Z(\zeta)\over \d\zeta}
  +
  \left(
  {\ml^2\over\zeta^2}
  +
  \kappa^4\zeta^2
  \right)
  Z(\zeta)
  &=
  \varepsilon
  Z(\zeta)
  \label{rp}
  \\
  \frac{\kappa^4}{(m_1+m_2)^2}
  \frac{\d}{\d x}
  \left[
    x(1-x) \frac{\d X(x)}{\d x}
    \right]
  +
  \left(
  M^2
  -
  {m_1^2\over x}
  -
  {m_2^2\over 1-x}
  \right)
  X(x)
  &=
  \varepsilon
  X(x)
  \label{X}
  \,,
\end{align}
where $\varepsilon$ is an eigenvalue to be solved for.


\subsubsection{Solutions for $Z(\zeta)$}

Let us find $Z(\zeta)$ first.
\eq{rp} can be simplified with the substitution
$Z(\zeta)={\phi(\zeta)\over \sqrt{\zeta}}$,
giving:
\begin{eqnarray}
  \left(
  -
  {\d^2\over \d \zeta^2}
  -
  {1-4\ml^2\over 4\zeta^2}
  +
  \kappa^4 \zeta^2
  \right)
  \phi(\zeta)
  =
  \varepsilon
  \phi(\zeta)
  \,.
\end{eqnarray}
This is equivalent to Eq.~(2.45) of Ref.~\cite{Brodsky:2014yha}
with $U(\zeta) = \kappa^4\zeta^2$,
and thus to the same reference's Eq.~(5.5) with $J=1$.
The solution---normalized to obey Eq.~(\ref{eqn:norm:Z})---is
therefore given by their Eq.~(5.6) with $J=1$:
\begin{eqnarray}
  \label{eqn:Z}
  Z_{\nt,\ml}(\zeta)
  =
  {\phi_{\nt,\ml}(\zeta)\over \sqrt{\zeta}}
  =
  \kappa
  \sqrt{2\,\nt!\over(\nt+\ml)!}
  (\kappa\zeta)^{\ml}
  e^{-\frac{1}{2}(\kappa\zeta)^2}
  L_{\nt}^{\ml}\big((\kappa\zeta)^2\big)
  \,,
\end{eqnarray}
where $L_{\nt}^{\ml}(z)$ are the associated Laguerre polynomials.
(Note that the $\lambda$ of Ref.~\cite{Brodsky:2014yha} is equal to our $\kappa^2$.)
The eigenvalues of this equation are:
\begin{eqnarray}
  \label{eqn:lambda}
  \varepsilon_{\nt,\ml}
  =
  2
  (2\nt + \ml + 1)
  \kappa^2
  \,,
\end{eqnarray}
where $\nt$ is a non-negative integer.
From the orthogonality and normalization properties
of the associated Laguerre polynomials, it follows that:
\begin{align}
  \int_0^\infty \d \zeta \,
  \zeta \,
  Z_{\nt,\ml}(\zeta)
  Z_{\nt',\ml}(\zeta)
  &=
  \delta_{\nt\nt'}
  \,,
\end{align}
so solutions with different $\nt$ eigenvalues are orthogonal,
and the required normalization rule (\ref{eqn:norm:Z}) is satisfied.
Recall that orthogonality for different $\ml$ eigenvalues is
already enforced in Eq.~(\ref{eqn:ortho:phi}).


\subsubsection{Solutions for $X(x)$}

We next turn to $X(x)$, as given by \eq{X}.
This equation has been solved by LMZV~\cite{Li:2015zda},
but we walk the reader through the solution
as a pedagogical guide.
This differential equation can be transformed
into the equation for Jacobi polynomials
through a series of steps,
which we will follow below.
First,
it is easier to analyze a differential equation
with only dimensionless coefficients,
so we make the substitutions:
\begin{align}
  \begin{split}
    c_0
    &=
    \frac{(m_1+m_2)^2(M^2-\varepsilon)}{\kappa^4}
    \\
    c_1
    &=
    \frac{m_1^2(m_1+m_2)^2 }{\kappa^4}
    \\
    c_2
    &=
    \frac{m_2^2(m_1+m_2)^2}{\kappa^4}
  \end{split}
\end{align}
to obtain:
\begin{eqnarray}
  x(1-x)
  X''(x)
  +
  (1-2x)
  X'(x)
  +
  \left(c_0-{c_1\over x}-{c_2\over 1-x}\right)
  X(x)
  =
  0
  \label{xeq}
  \,.
\end{eqnarray}
Next, we eliminate the terms proportional to
$\frac{1}{x}$ and $\frac{1}{1-x}$
in order to obtain a differential equation
with only polynomial coefficients.
This can be accomplished with the substitution:
\begin{eqnarray}
  X(x)
  =
  \sqrt{ x^{\alpha} (1-x)^{\beta}}
  f(x)
  \,,
  \label{Xeq}
\end{eqnarray}
where:
\begin{align}
  \label{eqn:ab}
  \begin{split}
    \alpha
    &=
    2
    \sqrt{c_1}
    =
    \frac{2 (m_1+m_2) m_1}{\kappa^2}
    \\
    \beta
    &=
    2 \sqrt{c_2}
    =
    \frac{2 (m_1+m_2) m_2}{\kappa^2}
    \,.
  \end{split}
\end{align}
The differential equation now becomes:
\begin{align}
  x(1-x)
  f''(x)
  +
  \Big( (\alpha+1)(1-x) - (\beta+1) x \Big)
  f'(x)
  +
  \left( c_0 - \frac{1}{4} (\alpha+\beta)(2+\alpha+\beta) \right)
  f(x)
  =
  0
  \label{feq}
  \,.
\end{align}
Finally, making the substitution gives
$x = \frac{1}{2} \big(1 - z\big)$,
the standard differential equation for Jacobi polynomials
$P_{\nl}^{(\alpha,\beta)}(z)$:
\begin{eqnarray}
  (1-z^2)
  f''(z)
  +
  2
  \Big(
  \beta
  - \alpha
  - (\alpha + \beta + 2) z
  \Big)
  f'(z)
  +
  \left( c_0 - \frac{1}{4} (\alpha+\beta)(2+\alpha+\beta) \right)
  f(z)
  =
  0
  \,,
  \label{f1}
\end{eqnarray}
as given for instance in
Eq.~(22.6.1) of Abramowitz \& Stegun~\cite{AS72},
with the eigenvalue for $c_0$ parametrized by the non-negative integer $\nl$ via:
\begin{align}
  \label{eqn:mu0}
  c_0
  &=
  \nl^2
  +
  (\alpha + \beta + 1) \nl
  +
  \frac{1}{4} (\alpha+\beta)(\alpha + \beta + 2)
  \,.
\end{align}
The $x$ dependence of the wave function is accordingly given by:
\begin{eqnarray}
  X_{\nl}(x)
  =
  \sqrt{
    \frac{
      4\pi
      (2\nl + \alpha + \beta + 1)
      \nl! \,
      \Gamma(\nl+\alpha+\beta+1)
    }{
      \Gamma(\nl + \alpha + 1)
      \Gamma(\nl + \beta + 1)
    }
  }
  \,
  \sqrt{ x^{\alpha}(1-x)^{\beta} }
  P_{\nl}^{(\alpha,\beta)}(1-2x)
  \,.
  \label{eqn:X}
\end{eqnarray}
From the orthogonality and normalization properties of the Jacobi polynomials,
it follows that:
\begin{align}
  \int_0^1 \frac{\d x}{4\pi}
  X_{\nl}(x)
  X_{\nl'}(x)
  =
  \delta_{\nl\nl'}
  \,,
\end{align}
so results with different eigenvalues are orthogonal,
and the $x$ dependence is normalized as required by
Eq.~(\ref{eqn:norm:X}).


\subsubsection{Overall solutions}

From Eqs.~(\ref{eqn:Z}) and (\ref{eqn:X}),
we find the overall wave function to be:
\begin{multline}
  \label{wfn}
  \psi_{\nl\nt\ml}(x,\zeta,\phi)
  =
  \kappa^{1+\ml}
  \sqrt{
    \frac{
      2
      (2\nl + \alpha + \beta + 1)
      \nt!
      \nl! \,
      \Gamma(\nl+\alpha+\beta+1)
    }{
      (\nt+\ml)!
      \Gamma(\nl + \alpha + 1)
      \Gamma(\nl + \beta + 1)
    }
  }
  \\
  \times
  \sqrt{ x^{\alpha+2}(1-x)^{\beta+2} }
  \zeta^{\ml}
  P_{\nl}^{(\alpha,\beta)}(1-2x)
  L_{\nt}^{\ml}\big((\kappa \zeta)^2\big)
  \e^{-\frac{1}{2}(\kappa\zeta)^2}
  \e^{i\ml\phi}
  \,,
\end{multline}
where we retain holographic variables because the solutions
have a simpler form in these coordinates.
The solutions are orthonormal
with respect to the appropriate measure---%
see Eq.~(\ref{eqn:norm:holo})---as required:
\begin{align}
  \int_0^1 \frac{\d x}{4\pi x^2(1-x)^2}
  \int_0^\infty \d \zeta \, \zeta
  \int_0^{2\pi} \d \phi \,
  \psi^*_{\nl'\nt'\ml'}(x,\zeta,\phi)
  \psi_{\nl\nt\ml}(x,\zeta,\phi)
  =
  \delta_{\nt\nt'}
  \delta_{\nl\nl'}
  \delta_{\ml\ml'}
  \,.
\end{align}
Additionally, by combining Eqs.~(\ref{eqn:lambda}) and (\ref{eqn:mu0}),
we find the following mass spectrum:
\begin{align}
  M^2_{\nl\nt\ml}
  =
  \kappa^2
  \left\{
    4\nt + 2\ml + 2
    +
    \frac{\kappa^2}{(m_1+m_2)^2}
    \left[
      \nl^2
      +
      (\alpha+\beta+1) \nl
      +
      \frac{1}{4}
      (\alpha+\beta) (\alpha+\beta+2)
      \right]
    \right\}
  \label{Msq}
  \,,
\end{align}
where we recall that $(\nl,\nt,\ml)$ are the quantum numbers,
$\kappa$ is a parameter from the harmonic oscillator potential (\ref{eqn:ho}),
and $\alpha$ and $\beta$ are auxiliary parameters defined in Eq.~(\ref{eqn:ab}).


\subsection{Non-relativistic limit}
\label{sec:nr}

We next consider the non-relativistic limit of our solutions.
This is helpful for two reasons.
First, it serves as a sanity check on our solutions;
if they are physically sensible,
they should reproduce the expected non-relativistic limit.
Second, by examining the non-relativistic limit,
we can establish a correspondence between the parameters
and quantum numbers appearing in our potential and solution,
and the usual parameters and quantum numbers considered in
the non-relativistic case.

Reproducing the correct non-relativistic limit requires firstly
that the spectrum of Eq.~(\ref{Msq}) reduces to the non-relativistic formula
\begin{align}
  \label{eqn:ENR}
  E_{\mathrm{NR}}
  =
  \omega
  \left(
  \frac{3}{2}
  +
  n
  \right)
  \,,
\end{align}
and secondly that the wave functions in Eq.~(\ref{wfn})
produce the expected solution for the three-dimensional
non-relativistic harmonic oscillator
in cylindrical coordinates:
\begin{align}
  \label{eqn:wfn:NR}
  \begin{split}
    \psi^{(\mathrm{NR})}_{\nl \nt \ml}(k_z, r_\perp, \phi)
    &=
    \widetilde{Z}_{\nl}(k_z)
    R_{\nt \ml}(r_\perp)
    \frac{\e^{i\ml\phi}}{\sqrt{2\pi}}
    \\
    R_{\nt\ml}(r_\perp)
    &=
    \sqrt{
      \frac{\mu\omega\nt!}{(\nt+\ml)!}
    }
    \big(\sqrt{\mu\omega} r_\perp)^{\ml}
    L_{\nt}^{\ml}\big(\mu\omega r_\perp^2\big)
    \e^{-\frac{1}{2} \mu\omega r_\perp^2}
    \\
    \widetilde{Z}_{\nl}(k_z)
    &=
    \sqrt{
      \frac{1}{2^{\nl} \nl!}
      \sqrt{\frac{4\pi}{\mu\omega}}
    }
    H_{\nl}\left(-\frac{k_z}{\sqrt{\mu\omega}}\right)
    \e^{-\frac{1}{2} \frac{k_z^2}{\mu\omega}}
    \,,
  \end{split}
\end{align}
where $H_n$ is the $n$th-order Hermite polynomial,
$\mu = \frac{m_1 m_2}{m_1 + m_2}$ is the reduced mass,
and where the quantum numbers indexing these solutions are related
to the principal quantum number by:
\begin{align}
  \label{eqn:principal}
  n
  =
  \nl
  +
  2 \nt
  +
  \ml
  \,.
\end{align}
These solutions can be obtained in a straightforward manner
by breaking down the three-dimensional harmonic oscillator into
a two-dimensional oscillator in polar coordinates
and a one-dimensional oscillator in the $z$ direction,
and taking the Fourier transform of the latter.
We represent the longitudinal degree of freedom in momentum space
to make the matching onto the momentum fraction variable more straightforward.


\subsubsection{Limit of the energy spectrum}

First, we check the spectrum of \eq{Msq}
in the non-relativistic limit.
This limit is implemented by considering $\kappa \ll m_1, m_2$.
Dropping $\mathcal{O}(\kappa^4)$ terms, we find:
\begin{align}
  M^2_{\nl\nt\ml}
  \approx
  (m_1 + m_2)^2
  +
  2
  \kappa^2
  \left(
  \nl + 2\nt + \ml + \frac{3}{2}
  \right)
  +
  \mathcal{O}(\kappa^4)
  \,.
\end{align}
Next, we note that, non-relativistically:
\begin{align}
  M^2
  =
  (m_1 + m_2 + E_{\mathrm{NR}})
  \approx
  (m_1 + m_2)^2
  +
  2 (m_1 + m_2) E_{\mathrm{NR}}
  +
  \mathcal{O}\big(E_{\mathrm{NR}}^2\big)
  \,.
\end{align}
The light front spectrum matches the non-relativistic spectrum
in the small-$\kappa$ limit, provided that:
\begin{align}
  \label{eqn:kappa:matching}
  \kappa^2
  =
  (m_1 + m_2) \omega
  \,.
\end{align}
Additionally, considering Eq.~(\ref{eqn:principal}),
the non-relativistic quantum numbers for cylindrical coordinate solutions
match exactly with the light front quantum numbers.


\subsubsection{Limit of the transverse coordinate dependence}

We next consider the non-relativistic limits of the wave functions.
The non-relativistic limit of the light front variables was explained
in Ref.~\cite{Miller:2009sg}.
In particular, $x$ is related to $k_z$ by:
\begin{align}
  x
  \approx
  \frac{m_1 + k_z}{m_1+m_2}
  \,,
  \qquad
  1-x
  \approx
  \frac{m_2 - k_z}{m_1+m_2}
  \,.
\end{align}
When taking the non-relativistic limit of
the $r_\perp$ dependence in $Z(\zeta)$,
we eliminate the coupling between the $r_\perp$ and $k_z$
variables by using the $k_z \ll m_1, m_2$ limit
in Eq.~(\ref{eqn:zeta}):
\begin{align}
  \zeta^2
  =
  x(1-x)
  r_\perp^2
  & \approx
  \frac{\mu r_\perp^2}{m_1+m_2}
  +
  \mathcal{O}\left( \left(\tfrac{k_z}{(m_1+m_2)}\right)^2 \right)
  \,.
\end{align}
Plugging this expression for $\zeta^2$ into Eq.~(\ref{eqn:Z})
gives exactly the expected non-relativistic wave function
$R_{\nt\ml}(r_\perp)$---as in Eq.~(\ref{eqn:wfn:NR})---provided that:
\begin{align*}
  \frac{\mu \kappa^2}{m_1+m_2}
  =
  \mu \omega
  \,.
\end{align*}
This condition is satisfied by Eq.~(\ref{eqn:kappa:matching}),
so the non-relativistic limit of the $r_\perp$ dependence is correct.


\subsubsection{Limit of the longitudinal momentum dependence}

Showing that the $x$ dependence has the correct non-relativistic limit is trickier.
First of all, the $k_z$ dependence is normalized as
\begin{align}
  \int \frac{\d k_z}{2\pi}
  \big| \widetilde{Z}(k_z) \big|^2
  =
  1
  \,,
\end{align}
so noting that $\d k_z = (m_1+m_2) \d x$
and comparing to Eq.~(\ref{eqn:norm:X}),
we require:
\begin{align}
  \label{eqn:correspondence:X}
  \frac{ X(x) }{\sqrt{2(m_1+m_2)}}
  \xrightarrow[\kappa,k_z \ll m_1, m2]{}
  \widetilde{Z}(k_z)
  \,.
\end{align}
We shall show, up to a normalization factor,
that this correspondence holds.
More specifically, we will show that when
$k_z, \kappa \ll m_1, m_2$, we have:
\begin{align}
  \label{eqn:x:matching}
  \begin{split}
    \sqrt{x^\alpha (1-x)^\beta}
    &
    \xrightarrow[\kappa,k_z \ll m_1, m2]{}
    \sqrt{ \frac{m_1^\alpha m_2^\beta}{(m_1+m_2)^{\alpha+\beta}} }
    \e^{-\frac{1}{2} \frac{k_z^2}{\mu\omega}}
    \\
    P_{\nl}^{(\alpha,\beta)}(1-2x)
    &
    \xrightarrow[\kappa,k_z \ll m_1, m2]{}
    \left(\frac{\sqrt{m_1m_2}}{\kappa}\right)^{\nl}
    H_{\nl}\left(-\frac{k_z}{\sqrt{\mu\omega}}\right)
    \,.
  \end{split}
\end{align}

We prove the first equation of (\ref{eqn:x:matching})
by expanding both sides to order $k_z^2$.
To this end, we first consider the expansions:
\begin{align}
  \begin{split}
    x^\alpha
    & \approx
    \frac{m_1^\alpha}{(m_1+m_2)^\alpha}
    \left(
    1
    +
    \alpha
    \frac{k_z}{m_1}
    +
    \frac{\alpha(\alpha-1)}{2}
    \frac{k_z^2}{m_1^2}
    +
    \ldots
    \right)
    \\
    (1-x)^\beta
    & \approx
    \frac{m_2^\beta}{(m_1+m_2)^\beta}
    \left(
    1
    -
    \beta
    \frac{k_z}{m_2}
    +
    \frac{\beta(\beta-1)}{2}
    \frac{k_z^2}{m_2^2}
    +
    \ldots
    \right)
    \,.
  \end{split}
\end{align}
By plugging in the formulas for $\alpha$ and $\beta$
from Eq.~(\ref{eqn:ab}),
and using the matching relation (\ref{eqn:kappa:matching}), we find:
\begin{align}
  \begin{split}
    x^\alpha
    & \approx
    \frac{m_1^\alpha}{(m_1+m_2)^\alpha}
    \left(
    1
    +
    \frac{2 k_z}{\omega}
    +
    \frac{(2 m_1 - \omega) k_z^2}{m_1 \omega^2}
    +
    \ldots
    \right)
    \\
    (1-x)^\beta
    & \approx
    \frac{m_2^\beta}{(m_1+m_2)^\beta}
    \left(
    1
    -
    \frac{2 k_z}{\omega}
    +
    \frac{(2 m_2 - \omega) k_z^2}{m_2 \omega^2}
    +
    \ldots
    \right)
    \,.
  \end{split}
\end{align}
Combining these and taking the square root gives:
\begin{align}
  \sqrt{
    x^\alpha (1-x)^\beta
  }
  \approx
  \sqrt{
    \frac{m_1^\alpha m_2^\beta}{(m_1+m_2)^{\alpha+\beta}}
  }
  \left(
  1
  -
  \frac{k_z^2}{2\mu\omega}
  +
  \ldots
  \right)
  \,.
\end{align}
Up to a normalization factor, this coincides with the order-$k_z^2$
expansion of $\e^{-\frac{1}{2}\frac{k_z^2}{\mu\omega}}$.

We prove the second equation of (\ref{eqn:x:matching})
using the framework developed by L\'opez and Temme~\cite{lopez1999approximations}
for approximating orthogonal polynomials via Hermite polynomials.
In particular, Eq.~(6.3) of Ref.~\cite{lopez1999approximations}
gives an expansion of Jacobi polynomials in terms of Hermite polynomials\footnote{
  Some variables from Ref.~\cite{lopez1999approximations}
  were changed to avoid notation clashes.
}:
\begin{align}
  \label{eqn:lopez}
  P_n^{(\alpha,\beta)}(\chi)
  =
  B^{n/2}
  \sum_{k=0}^n
  \frac{c_k}{B^{k/2}}
  \frac{H_{n-k}(\eta)}{(n-k)!}
\end{align}
where
\begin{align}
  \begin{split}
    A
    &=
    \frac{1}{2}
    \Big(
    \alpha - \beta + (\alpha + \beta + 2) \chi
    \Big)
    \\
    B
    &=
    \frac{1}{8}
    \Big(
    \alpha + \beta + 4
    + 2(\beta - \alpha) \chi
    -
    (3\alpha + 3\beta + 8) \chi^2
    \Big)
    \\
    \eta
    &=
    \frac{A}{2\sqrt{B}}
    \,.
  \end{split}
\end{align}
The first few coefficients are $c_0 = 1$ and $c_1 = c_2=0$~\cite{lopez1999approximations}.
In this context,
\begin{align}
  \chi
  =
  1 - 2x
  & \approx
  \frac{m_2 - m_1}{m_1 + m_2}
  -
  \frac{2 k_z}{m_1 + m_2}
  \,,
\end{align}
and in the $\kappa,k_z \ll m_1,m_2$ limit, we find:
\begin{align}
  A
  \approx
  -\frac{2 (m_1+m_2) k_z}{\kappa^2}
  \,, \qquad
  B
  \approx
  \frac{\mu(m_1+m_2)}{\kappa^2}
  \,, \qquad
  \eta
  \approx
  - \frac{k_z}{\sqrt{\mu\omega}}
  \,.
\end{align}
Since $B^{-k/2}$ falls as order $\kappa/m$,
and since the first non-zero correction to the leading $k=0$ term is $k=3$,
the expansion (\ref{eqn:lopez}) can be truncated at $k=0$ for $\kappa \ll m_1,m_2$.
This truncation gives the second equation of (\ref{eqn:x:matching}),
thus proving that the $x$ dependence of our solution
produces the correct non-relativistic limit.

\subsubsection{Numerical comparisons of light front and non-relativistic solutions}

\begin{figure}
  \includegraphics[width=\textwidth]{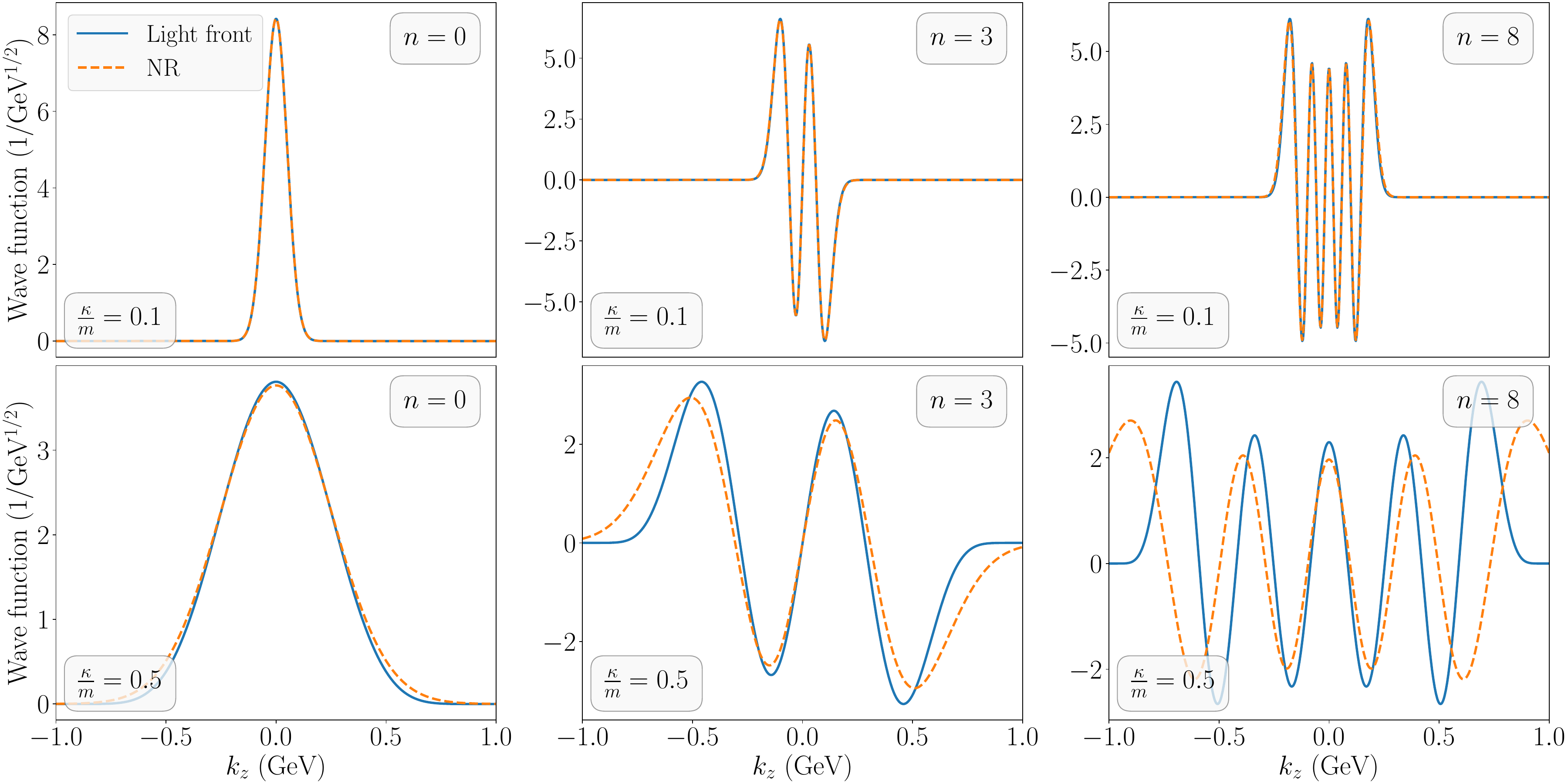}
  \caption{
    A comparison of the longitudinal momentum dependence of the
    light front (solid blue lines)
    and non-relativistic (dashed orange lines)
    solutions to the harmonic oscillator.
    More specifically, the left- and right-hand sides of
    Eq.~(\ref{eqn:correspondence:X}) are compared in this plot.
    We use $m_1 = m_2 \equiv m = 1$~GeV.
  }
  \label{fig:nr}
\end{figure}

We lastly consider some numerical examples,
both to further demonstrate that the $x$ dependence of our solutions reproduces
the correct non-relativistic limit,
and to give a rough idea when relativistic corrections become important.
The key condition behind the non-relativistic limit is that $\kappa \ll m_1, m_2$.
We consider the equal mass case
with $m_1 = m_2 \equiv m = 1$~GeV
for a simple concrete illustration.

In Fig.~\ref{fig:nr},
we give comparisons for $\kappa = 0.1$~GeV and $\kappa = 0.5$~GeV.
The top row has $\kappa/m = 0.1$,
for which the light front solutions are visually indistinguishable
from the non-relativistic solutions.
The approximation of Jacobi polynomials using a single Hermite polynomial
is well-justified for these parameters, as $\alpha = \beta \approx 400$.
The bottom row, however, has $\kappa=0.5$ and shows stark differences
between the light front and non-relativistic solutions---which become
more extreme at larger $n$.
For these parameters, $\alpha = \beta \approx 16$,
and the Jacobi polynomials for this $(\alpha,\beta)$ pair are not
well-approximated by a single Hermite polynomial.

To give a more concrete and intuitive idea of the physical meanings of
these parameter values,
we can use Eq.~(\ref{eqn:kappa:matching}) to obtain the corresponding $\omega$ values.
For $\kappa/m = 0.1$ and $m=1$~GeV, we find
$\omega \approx 7.6$~ZHz
(i.e., $7.6 \times 10^{21}$~Hz),
while for $\kappa/m = 0.5$ we find
$\omega \approx 190$~ZHz.
These frequencies are absurdly high by the standards of molecular physics,
meaning the non-relativistic limit is very well-justified in molecules.
For bound systems of quarks, however,
such frequencies are natural;
a rough phenomenological estimate finds that the up and down quarks
orbit the proton with orbital frequencies of around
$125$~ZHz and $276$~ZHz, respectively~\cite{Freese:2023jcp}.
Moreover, rearranging Eq.~(\ref{eqn:kappa:matching}) slightly gives:
\begin{align}
  \omega
  =
  \left(\frac{\kappa}{2m}\right)^2
  2m
  \,,
\end{align}
meaning that for fixed $\kappa/m$,
the natural frequency of the oscillator scales linearly with the constituent mass.
For current quark masses of around 3~MeV (as an example),
we'd find $\kappa/m=0.1$ and $\kappa/m=0.5$ to correspond to natural oscillator
frequencies of about $22.8$~EHz and $571$~EHz, respectively---well less
than the orbital frequencies of up and down quarks in a proton.
This suggests that the non-relativistic limit is wholly inadequate
in hadronic systems,
and that a relativistic treatment in terms of the light front is necessary.

\begin{figure}
  \includegraphics[width=\textwidth]{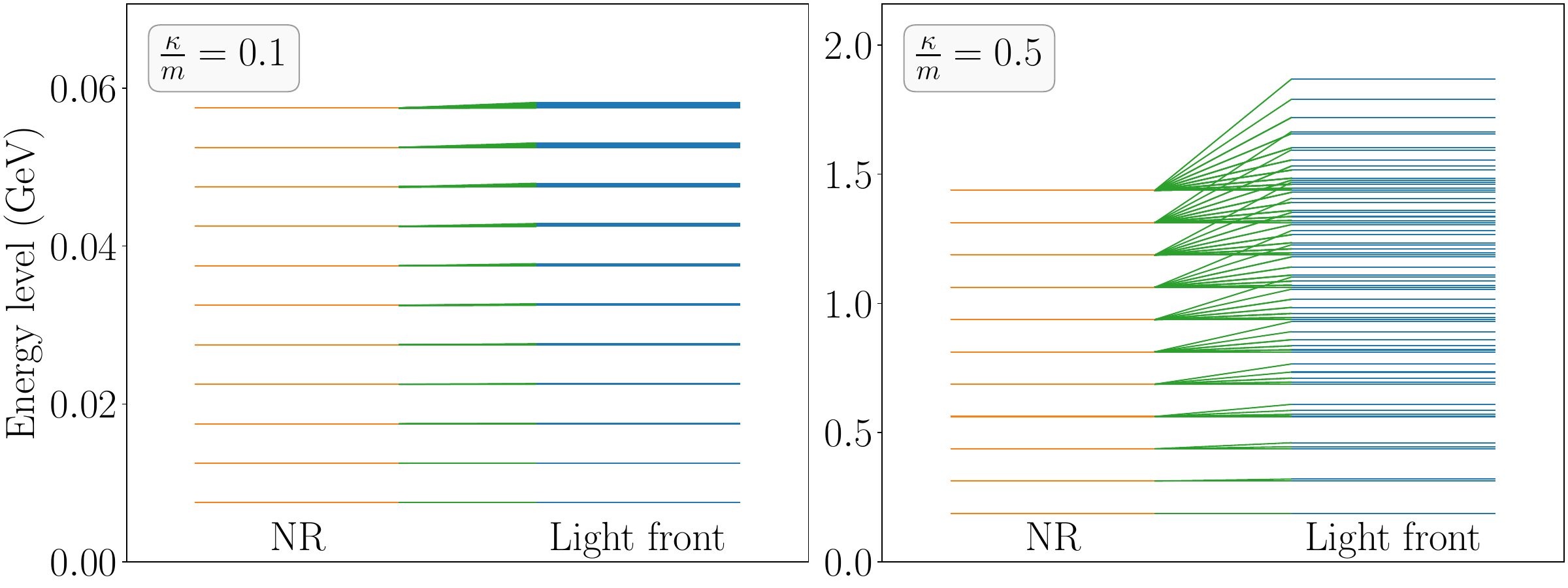}
  \caption{
    The energy levels for the first eleven
    principal quantum numbers ($0 \leq n \leq 10$)
    of the harmonic oscillator.
    These plots compare the non-relativistic results for the energy levels
    as defined in Eq.~(\ref{eqn:ENR})
    (orange lines)
    to the light front energy levels
    as defined in Eq.~(\ref{eqn:ELF})
    (blue lines).
    We also draws lines connecting the light front energy levels
    to the corresponding non-relativistic levels
    (green lines).
    Both panels use
    $m_1 = m_2 \equiv m = 1$~GeV.
  }
  \label{fig:energy}
\end{figure}

Finally, we compare the non-relativistic energy levels to the corresponding
light front energy levels.
In this context, we use Eq.~(\ref{eqn:ENR}) for the non-relativistic energies,
and subtract off the rest mass from the light front results.
In effect, we use:
\begin{align}
  \label{eqn:ELF}
  E_{\mathrm{LF}}
  \equiv
  \frac{1}{4m}
  \Big( M^2 - (2m)^2 \Big)
  \,.
\end{align}
In the non-relativistic limit, $E_{\mathrm{LF}} \rightarrow E_{\mathrm{NR}}$.
We again use $m_1 = m_2 \equiv m = 1$~GeV
for a simple concrete illustration.
The results for principal quantum numbers
$0 \leq n \leq 10$ are shown in Fig.~\ref{fig:energy}.
The non-relativistic result depends only on the principal quantum number
$n = \nl + 2\nt + \ml$,
while the light front result contains terms that are quadratic in $\nl$
and accordingly splits the energy levels.
For $\kappa/m=0.1$, the splitting is minimal and the non-relativistic formula
provides an accurate solution.
For $\kappa=0.5$, the splitting becomes extreme---especially at higher $n$---and
energy levels from different $n$ even begin crossing.
The non-relativistic formula (\ref{eqn:ENR}) thus becomes inaccurate at
moderate $\kappa/m$,
and the relativistic formula (\ref{Msq}) becomes necessary.


\section{Summary and Outlook}
\label{sec:end}

Light front quantum mechanics uniquely permits the formulation of
a bound state wave equation for composite relativistic systems.
Moreover, the wave equation and its solutions are boost-invariant,
meaning the wave function is a genuine description of internal structure.
The most crucial component of this outcome is that separation of variables
can be achieved between barycentric and internal degrees of freedom,
which in turn is a consequence of the light front having a kinematic subgroup
that includes boosts.
This outcome cannot be achieved in instant form coordinates;
for example, the free instant form energy for two particles,
\begin{align*}
  E_1 + E_2
  =
  \sqrt{m_1^2 + \bm{p}_1^2} + \sqrt{m_2^2 + \bm{p}_2^2}
  \,,
\end{align*}
cannot be rewritten as a sum of terms depending both on a relative momentum
and the total momentum $\bm{P} = \bm{p}_1 + \bm{p}_2$.
On the other hand, the free light front energy
\begin{align*}
  P_1^-
  +
  P_2^-
  =
  \frac{m_1^2 + \bm{p}_{1\perp}^2}{2p_1^+}
  +
  \frac{m_2^2 + \bm{p}_{2\perp}^2}{2p_2^+}
  =
  \frac{1}{2P^+}
  \left(
  \bm{P}_\perp^2
  +
  \frac{m_1^2}{x}
  +
  \frac{m_2^2}{1-x}
  +
  \frac{\bm{k}_\perp^2}{x(1-x)}
  \right)
\end{align*}
does,
with
$\bm{k}_\perp = (1-x) \bm{p}_{1\perp} - x\bm{p}_{2\perp}$
and $x = \frac{p_1^+}{P^+}$.
The interpretation of $\bm{k}_\perp$ and $x$ as genuine internal
variables is bolstered by their invariance under both boosts
and translations;
see Sec.~\ref{sec:separation} and the Appendix for technical details.

In this work, we utilized this unique capability of the light front
to formulate and solve a relativistic wave equation
for a bound system of two particles.
We showed that the
relativistic harmonic oscillator potential proposed by
Li, Maris, Zhao and Vary~\cite{Li:2015zda,Li:2021jqb}
appears naturally when considering the Miller-Brodsky variable $\tilde{z}$,
which provides a boost-invariant description of longitudinal spatial separation.
We then posed and solved a relativistic wave equation with this potential,
finding exact, closed-form solutions.
We showed in Sec.~\ref{sec:nr} that these reduce to the known solutions
from the textbook harmonic oscillator problem
in the non-relativistic limit.
We find relativistic effects to be significant when the harmonic oscillator
constant $\kappa$ is large compared to either of the constituent masses.

The formalism presented in Sec.~\ref{sec:lf} is quite general,
and can be generalized further to systems with any number of particles.
It may be difficult to find closed-form solutions like those of
Sec.~\ref{sec:harmonic}.
However, standard numerical methods and approximation methods---such
as variational methods, expansions in harmonic oscillator solutions,
perturbation theory, etc.---can be used for other potentials.
Indeed, even in non-relativistic quantum mechanics,
such numerical methods are used in all but a handful of specific
two-body potentials.

Future avenues of research will include the formulation and
numerical solution of wave equations with other potentials.
A light front formulation of the Cornell potential could be especially
pertinent to describing light meson structure.
Besides this, the extension of Sec.~\ref{sec:separation} to three-body
systems merits thorough investigation.

Lastly, besides extension to other potentials and many-body systems,
it would be interesting to calculate generalized parton
distributions~\cite{Muller:1994ses,Ji:1996nm,Radyushkin:1997ki},
electromagnetic and mechanical form factors
(and the light front densities entailed by them),
and to test polynomiality~\cite{Ji:1998pc} in this formalism.

\begin{acknowledgments}
  We warmly acknowledge helpful discussions with Dimitriy Kim,
  Cédric Lorcé and Philip Mannheim.
  We thank Eric Fu for a careful reading of the manuscript
  and for catching a typo in an intermediate step.
  AF was supported by
  the Center for Nuclear Femtography,
  operated by the Southeastern Universities Research Association
  in Washington, D.C.\ under an appropriation from the Commonwealth of Virginia.
  GAM is partially supported by the U.S.\ Department of Energy,
  Office of Science, Office of Nuclear Physics,
  under contract number DE-SC0026252.
  This material is based upon work supported by the U.S.\ Department of Energy,
  Office of Science, Office of Nuclear Physics under
  Contract No.\ 89243126CSC000213.
  The research in this work received inspiration from the goals of the
  Quark Gluon Tomography Topical Collaboration
  of the U.S.\ Department of Energy.
\end{acknowledgments}

\appendix


\section{The Poincar\'e algebra in light front coordinates}
\label{sec:poincare}

This appendix reviews the algebra of generators of the Poincar\'e group
when cast into light front coordinates.
The material herein is already well-known
(see e.g.\ Refs.~\cite{Dirac:1949cp,Susskind:1967rg,Kogut:1969xa,Soper:1971sr,Burkardt:2002hr,Lorce:2018zpf}),
but is reproduced here to make the work self-contained.
The Poincar\'e algebra consists of ten generators of the Poincar\'e group,
which obey the following commutation relations:
\begin{align}
  \begin{split}
    \big[ M^{\mu\nu}, M^{\rho\sigma} \big]
    &=
    i \big(
    g^{\mu\sigma}
    M^{\nu\rho}
    +
    g^{\nu\rho}
    M^{\mu\sigma}
    -
    g^{\mu\rho}
    M^{\nu\sigma}
    -
    g^{\nu\sigma}
    M^{\mu\rho}
    \big)
    \\
    \big[ M^{\mu\nu}, P^\rho \big]
    &=
    i \big(
    g^{\nu\rho}
    P^\mu
    -
    g^{\mu\rho}
    P^\nu
    \big)
    \\
    \big[ P^\mu, P^\nu \big]
    &=
    0
    \,.
  \end{split}
\end{align}
Here, $M^{\mu\nu} = -M^{\nu\mu}$ are the six Lorentz transformation generators,
and $P^\mu$ are the translation generators.
The translation generators in light front coordinates are simply enough:
\begin{align}
  P^\mu
  =
  (P^+; \bm{P}_\perp; P^-)
  \,.
\end{align}
The Lorentz transformation generators are sorted into boosts and rotations,
which in light front coordinates can be written:
\begin{align}
  \begin{split}
    \mathbf{Boosts}
    \qquad &: \qquad
    (B_1, B_2, K_3)
    =
    (M^{+1}, M^{+2}, -M^{+-})
    \\
    \mathbf{Rotations}
    \qquad &: \qquad
    (A_1, A_2, J_3)
    =
    (M^{2-}, M^{-1}, M^{12})
    \,.
  \end{split}
\end{align}
The labels $K_3$ and $J_3$ are chosen for longitudinal boosts
and rotations around the longitudinal axis because these coincide with
the instant form generators.
The other boosts $B_a$ and rotations $A_a$ do not coincide with instant form
generators, so we have given them different labels.
The rotations around the transverse axes can be written
\begin{align}
  A_a
  =
  -
  \epsilon_{ab3}
  M^{-b}
  \,,
\end{align}
which is helpful when evaluating commutators.

We find the following explicit commutation rules:
\begin{align}
  \label{eqn:poincare:lf}
  \begin{split}
    [B_a, B_b]
    &=
    0
    \\
    [B_a, K_3]
    &=
    i B_a
    \\
    [A_a, A_b]
    &=
    0
    \\
    [J_3, A_a]
    &=
    i\epsilon_{ab3} A_b
    \\
    [J_3, B_a]
    &=
    i\epsilon_{ab3} B_b
    \\
    [K_3, A_a]
    &=
    i A_a
    \\
    [B_a, A_b]
    &=
    i \delta_{ab} J_3 - i \epsilon_{ab3} K_3
    \\
    [B_a, P_\perp^b]
    &=
    - i\delta_{ab} P^+
    \\
    [B_a, P^+]
    &=
    0
    \\
    [B_a, P^-]
    &=
    -i P_\perp^a
    \\
    [K_3, P_\perp^a]
    &=
    0
    \\
    [K_3, P^+]
    &=
    -i P^+
    \\
    [K_3, P^-]
    &=
    i P^-
    \\
    [A_a, P_\perp^b]
    &=
    - i \epsilon_{ab3} P^-
    \\
    [A_a, P^+]
    &=
    i\epsilon_{ab3} P_\perp^b
    \\
    [A_a, P^-]
    &=
    0
    \\
    [J_3, P_\perp^a]
    &=
    i \epsilon_{ab3} P_\perp^b
    \\
    [J_3, P^+]
    &=
    0
    \\
    [J_3, P^-]
    &=
    0
    \,.
  \end{split}
\end{align}

\begin{table}
  \renewcommand{\arraystretch}{2.0}
  \caption{
    Lists of generators in important subalgebras
    of the Poincar\'e algebra
  }
  \begin{tabular}{cccccc}
    \toprule
    ~~~Galilei subalgebra~~~ &
    ~~~$P^+$~~~ &
    ~~~$\bm{P}_\perp$~~~ &
    ~~~$P^-$~~~ &
    ~~~$\bm{B}$~~~ &
    ~~~$J_3$~~~ \\
    ~~~Kinematic subalgebra~~~ &
    ~~~$P^+$~~~ &
    ~~~$\bm{P}_\perp$~~~ &
    ~~~$K_3$~~~ &
    ~~~$\bm{B}$~~~ &
    ~~~$J_3$~~~ \\
    \bottomrule
  \end{tabular}
  \label{tab:subgroups}
\end{table}

The Poincar\'e group contains two important subgroups
(and accordingly, the Poincar\'e algebra two important subalgebras)
for light front physics.
These are the Galilei subgroup and the kinematic subgroup.
The generators of each subgroup are summarized in Table~\ref{tab:subgroups}.
Both subgroups have seven generators, six of which coincide.

The Galilei subgroup consists of transformations that leave $P^+$ invariant,
and is isomorphic to the symmetry group of two-dimensional non-relativistic
mechanics.
This subgroup includes transverse boosts, rotation around the $z$ axis,
and all four components of the four-momentum.
It excludes longitudinal boosts because these alter $P^+$.
The main utility of this subgroup is that it explains many of
the formal similarities between light front physics
and non-relativistic physics.

The kinematic subgroup consists of transformations that leave the
condition $x^+=0$ invariant~\cite{Dirac:1949cp}.
This subgroup includes transverse boosts, rotation around the $z$ axis,
boosts along the $z$ axis, and three components $(P^+, \bm{P}_\perp)$
of the four-momentum.
The time translation generator $P^-$ is excluded because this evolves
$x^+$, and accordingly does not retain the condition $x^+=0$.
The main utility of this group is that,
for multi-particle systems,
each of its generators can cleanly be broken into pieces that act
separately on each particle.


\subsection{Position operators}

Light front position operators can helpfully be defined in terms of
generators of the Poincar\'e group.
Transverse and longitudinal position operators can be defined as follows:
\begin{align}
  \begin{split}
    X_\perp^a
    &=
    \frac{-B_a}{P^+}
    \\
    X^-
    &=
    \frac{1}{2}
    \left(
    \frac{1}{P^+}
    K_3
    +
    K_3
    \frac{1}{P^+}
    \right)
    \,.
  \end{split}
\end{align}
These operators are Hermitian and obey the expected canonical
commutation relations:
\begin{align}
  \begin{split}
    &
    [X_\perp^a, X_\perp^b]
    =
    [X_\perp^a, X^-]
    =
    [X_\perp^a, P^+]
    =
    [X^-, P_\perp^a]
    =
    0
    \\
    &
    [X_\perp^a, X_\perp^b]
    =
    i \delta_{ab}
    \\
    &
    [X^-, P^+]
    =
    - i
    \,.
  \end{split}
\end{align}
It is also helpful to note that the position operators are built
entirely from generators of the kinematic subgroup---see
Table~\ref{tab:subgroups}---so it is possible to construct position operators
for individual particles,
each of which individually obey the canonical commutation relations.

The transverse position operator is invariant under boosts:
\begin{align}
  \begin{split}
    [B_a, X_\perp^b]
    &=
    0
    \\
    [K_3, X_\perp^b]
    &=
    0
  \end{split}
\end{align}
while the longitudinal position operator is not:
\begin{align}
  \begin{split}
    [B_a, X^-]
    &=
    - i X_\perp^a
    \\
    [K_3, X^-]
    &=
    i X^-
    \,.
  \end{split}
\end{align}


\subsection{Miller-Brodsky variable}

We will here prove that the Miller-Brodsky variable,
as defined in Eq.~(\ref{eqn:ztilde}),
is boost-invariant, is independent of the barycentric position and momentum,
and is canonically conjugate to the momentum fraction $x$.

We consider boost invariance first.
Under infinitesimal longitudinal boosts,
the Miller-Brodsky variable transforms as:
\begin{align}
  [K_3, \tilde{z}]
  =
  [K_3, P^+] (x_2^- - x_1^-)
  +
  P^+ [K_3, x_2^- - x_1^-]
  +
  [K_3, \bm{P}_\perp\cdot\bm{r}_\perp]
  =
  - i P^+
  (x_2^- - x_1^-)
  +
  P^+ \Big( i(x_2^- - x_1^-) \Big)
  +
  0
  =
  0
  \,,
\end{align}
and is thus invariant.
Under infinitesimal transverse boosts, it transforms as:
\begin{align}
  [B_a, \tilde{z}]
  =
  P^+ [B_a, x_2^- - x_1^- ]
  +
  [B_a, P_\perp^b] r_\perp^b
  =
  P^+ \Big( i r_\perp^a \Big)
  +
  -i\delta_{ab} r_\perp^b
  =
  0
  \,,
\end{align}
and is thus invariant.
Therefore, $\tilde{z}$ is invariant under all boosts.

We next show that $\tilde{z}$ is canonically conjugate to $x$.
We have:
\begin{multline}
  [\tilde{z}, x]
  =
  P^+
  \left[
    x_2^- - x_1^-,
    \frac{p^+_1}{P^+}
    \right]
  +
  \left[
    x_2^- - x_1^-,
    \bm{P}_\perp\cdot\bm{r}_\perp
    \right]
  \\
  =
  P^+
  \left(
  -
  p_1^+
  [x_2^-, P^+]
  \frac{1}{(P^+)^2}
  -
  [x_1^-, p_1^+]
  \frac{1}{P^+}
  +
  p_1^+
  [x_1^-, P^+]
  \frac{1}{(P^+)^2}
  \right)
  +
  0
  =
  i
  \,.
\end{multline}
This is as required.

Finally, we consider independence from barycentric variables.
Independence is established if $\tilde{z}$ commutes
with both $\bm{R}_\perp$ and $\bm{P}_\perp$.
The proof will exploit our finding that $\tilde{z}$ and $x$
are canonically conjugate.
Considering $\bm{P}_\perp$ first:
\begin{align}
  [\tilde{z}, P_\perp^a]
  =
  [P^+(x_2^--x_1^-), P_\perp^a]
  +
  [\bm{P}_\perp\cdot\bm{r}_\perp, P_\perp^a]
  =
  0
  \,,
\end{align}
which is trivially zero.
Next, we consider $\bm{R}_\perp = x \bm{x}_{1\perp} + (1-x) \bm{x}_{2\perp}$:
\begin{align}
  [\tilde{z}, R_\perp^a]
  =
  [\tilde{z}, x] x_{1\perp}^a
  +
  [\tilde{z}, (1-x)] x_{b\perp}^a
  +
  [P_\perp^b, R_\perp^a] r_\perp^b
  =
  i x_{1\perp}^a
  - i x_{2\perp}^a
  - i \delta_{ab} r_\perp^b
  =
  0
  \,.
\end{align}
This proves $\tilde{z}$ has all the necessary properties.


\bibliography{references.bib}

\end{document}